\RequirePackage{ifpdf}
\ifpdf 
\documentclass[pdftex]{sigma}
\else
\documentclass{sigma}
\fi

\begin{document}

\renewcommand{\thefootnote}{\fnsymbol{footnote}}
\newcommand{\Tr}{\textrm{Tr}}
\newcommand{\Lie}{\operatorname{Lie}}
\newcommand{\Pic}{\operatorname{Pic}}
\newcommand{\C}{\operatorname{\mathbb{C}}}
\newcommand{\R}{\operatorname{\mathbb{R}}}
\newcommand{\Q}{\operatorname{\mathbb{Q}}}
\newcommand{\Z}{\operatorname{\mathbb{Z}}}
\newcommand{\N}{\operatorname{\mathbb{N}}}
\newcommand{\I}{\operatorname{\mathbb{I}}}
\newcommand{\cM}{\operatorname{\mathcal{M}}}
\newcommand{\Hom}{\operatorname{\mathrm{Hom}}}

\newcommand{\transpose}{{}^t}
\newcommand{\Mbar}{M^{\bullet}}

\allowdisplaybreaks
\renewcommand{\PaperNumber}{007}

\FirstPageHeading

\renewcommand{\thefootnote}{$\star$}

\ShortArticleName{Multi-Hamiltonian Structures on Beauville's Integrable System
and Its Variant}

\ArticleName{Multi-Hamiltonian Structures\\ on Beauville's Integrable System
and Its Variant\footnote{This paper is a contribution 
to the Vadim Kuznetsov Memorial Issue ``Integrable Systems and Related Topics''.
The full collection is available at 
\href{http://www.emis.de/journals/SIGMA/kuznetsov.html}{http://www.emis.de/journals/SIGMA/kuznetsov.html}}}

\Author{Rei INOUE~$^\dag$ and Yukiko KONISHI~$^\ddag$}
\AuthorNameForHeading{R. Inoue and Y. Konishi}

\Address{$^\dag$~Department of Physics, Graduate School of Science,
The University of Tokyo,\\
$\phantom{^\dag}$~7-3-1 Hongo, Bunkyo, Tokyo 113-0033, Japan} 

\EmailD{\href{mailto:reiiy@monet.phys.s.u-tokyo.ac.jp}{reiiy@monet.phys.s.u-tokyo.ac.jp}} 

\Address{$^\ddag$~Graduate School of Mathematical Sciences, The University of Tokyo,\\
$\phantom{^\ddag}$~3-8-1 Komaba, Meguro, Tokyo 153-8914, Japan}
\EmailD{\href{mailto:konishi@ms.u-tokyo.ac.jp}{konishi@ms.u-tokyo.ac.jp}}

\ArticleDates{Received October 24, 2006, in f\/inal form December 29, 2006; 
 Published online January 08, 2007}

\Abstract{We study Beauville's completely integrable 
  system and its variant from a viewpoint of multi-Hamiltonian structures.
  We also relate our result to the previously known Poisson structures
  on the Mumford system and the even Mumford system.}


\Classification{37J35; 14H70}

\section{Introduction}

Let $r\geq2$ and $d\geq 1$ be integers.
In \cite{Beauville90}, 
Beauville constructed  an algebraically 
completely integrable Hamiltonian system
on the space 
of the gauge equivalence classes of 
$r\times r$ polynomial matrices whose degree is less than or equal to $d$.
This system is a generalization of the Mumford system \cite{Mumford-Book},
and the general level set is isomorphic to 
the complement of the theta divisor in
the Jacobian variety of the spectral curve. 
Employing Beauville's approach,
a variant is constructed in \cite{IKY06},  
which gives a generalization of the even Mumford system 
introduced by Vanhaecke \cite{Van92,Van1638}.
The remarkable dif\/ference with Beauville's  system
is that the general level set is isomorphic 
to the complement of the intersection of $r$ translates of the 
theta divisor. 
We call these systems the Beauville system and 
the Beauville--Vanhaecke (BV) system in this paper.

The Poisson structures of these systems depend on 
a monic polynomial of degree $d+2$ having only simple roots.
Our f\/irst aim in this article is to understand
them in terms of the multi-Hamiltonian structure
(see \cite[\S~4 and \S~12]{ReySeme94}).
This is achieved by extending them 
to those parameterized by a polynomial $\phi(x)$ of degree at most $d+2$;
$\phi(x)$ need not be  monic nor  have simple roots,
and $\deg \phi(x)$ can be less than $d+2$
(Propositions \ref{prop:Poisson-Beauville} and \ref{lem:Poisson-BV}).
It then turns out that the Poisson structures corresponding to
$\phi(x)=1,x,\ldots,x^{d+2}$ give the integrable multi-Hamiltonian system
(Theorems \ref{thm:m-Ham-Beauville} and \ref{thm:m-Ham-BV}).

Since the Beauville system and the BV system are given  
by the quotient construction,
it is an important problem to construct their spaces of representatives
and to describe the vector f\/ields and the Poisson structures on them.
For the Beauville system, 
a space of representatives was constructed  by Donagi and Markman for a certain subspace \cite{DonagiMarkman96}.
The vector f\/ields on it was computed  by Fu \cite{Fu03} and 
the multi-Hamiltonian structure was constructed in \cite{PenVan98}
using the group-theoretic approach.
For the BV system, spaces of representatives were given in \cite{IKY06}
and the vector f\/ields for $r=2$ and $3$ were also given.

Our second aim is to study the family of Poisson structures 
on the space of representatives.
For the Beauville system, 
we introduce a new  space of representatives for a certain subspace 
of codimension one which includes Donagi--Markman's result
(Proposition \ref{prop:rep}).
In the simplest case of $r=2$
we write down the Poisson structures 
on this space and on the spaces of representatives for the BV system 
(Propositions \ref{prop:Poisson-B-2} and \ref{prop:Poisson-BV-2})
and compare them with previously known Poisson structures on
the Mumford system and the even Mumford system 
\cite{FernadesVanhae01,SmirnovNakayashiki00,PenVan98}.

This article is organized as follows.
In Sections~\ref{section:Beauville} and~\ref{section:BV},
we study Poisson structures 
and multi-Hamiltonian structures  
for the Beauville system and the BV system.
We also give  expressions of the Poisson structures
on the spaces of representatives in the case $r=2$
and compare them with those of the Mumford system and the even Mumford system.
Section~\ref{section:rep-Beauville} is devoted to the 
construction of  the new  space of the representatives for the  subset of 
the Beauville system. 

\renewcommand{\thefootnote}{\arabic{footnote}}
\setcounter{footnote}{0}

\section{Multi-Hamiltonian structure on the Beauville system}
\label{section:Beauville}

We f\/ix numbers $r \in \Z_{\geq 2}$ and $d \in \Z_{\geq 1}$.
Throughout this paper,  
we use the following notations:
$S_d\subset \mathbb{C}[x]$ denotes the set of polynomials of degree 
at most $d$.
$E_{ij}\in M_r(\C)$ is the matrix whose $(i,j)$-th entry is one and other entries are zero.
For a matrix $A(x)\in M_r(\mathbb{C}[x])$ with polynomial entries,
\begin{alignat*}{3} 
&      A_{ij}(x)\in\mathbb{C}[x]:&& \text{ the $(i,j)$-th entry of $A(x)$},&\\
&      A_{ij;k}\in \mathbb{C}:   && \text{ the coef\/f\/icient of $x^k$ in $A_{ij}(x)$},&\\
&      A_{k}\in M_r(\mathbb{C}):\qquad && \text{ the coef\/f\/icient of $x^k$ in $A(x)$}.&
\end{alignat*}

Let $W$ be a nonsingular algebraic variety.
\begin{definition}
A Poisson algebra structure on a sheaf of rings $\mathcal{F}$ on $W$ is a
morphism $\{\cdot,\cdot\}:\mathcal{F}\times \mathcal{F}\to\mathcal{F}$ satisfying
skew-symmetry, the Leibniz rule, and the Jacobi identity.
A Poisson structure on $W$ is a Poisson algebra structure 
on the structure sheaf $\mathcal{O}_W$.
\end{definition}

\subsection{The Beauville system}
Let $V(r,d)$ be the set
\[ 
  V(r,d)=\{P(x,y)=y^r+s_1(x)y^{r-1}+\cdots+ s_r(x) \in \mathbb{C}[x,y]
                      ~|~ s_i(x)\in S_{di}\}.
\]
For $P(x,y)\in V(r,d)$, let $C_{P}$ be the spectral curve 
obtained by taking the closure of the af\/f\/ine curve
$P(x,y)=0$ in the Hirzebruch surface 
$\mathbb{F}_d=\mathbb{P}(\mathcal{O}_{\mathbb{P}^1}\oplus\mathcal{O}_{\mathbb{P}^1}(d))$ of degree $d$.
Consider the set $M_r(S_d)$ of $r\times r$ matrices with entries in $S_d$
and let $\psi$ be  the map:
\begin{gather*}
\psi:M_r(S_d) \to V(r,d),\qquad 
  A(x)\mapsto \det (y\mathbb{I}_r-A(x)).
\end{gather*}
The group  $PGL_r(\mathbb{C})$ acts on $M_r(S_d)$ by conjugation:
\begin{gather*}
   PGL_r(\mathbb{C})\ni g: A(x)\mapsto g(A(x)) = g^{-1}A(x)g.
\end{gather*}
Def\/ine a subset $M_r(S_d)_{ir}$ of $M_r(S_d)$ as
\[
  M_r(S_d)_{ir}=\{A(x)\in M_r(S_d)~|~
                  \text{$C_{\psi(A(x))}$ is irreducible}\}.
\]
Note that the $PGL_r(\mathbb{C})$-action is free on $M_r(S_d)_{ir}$. 
Let $\eta: M_r(S_d)_{ir}\to M_r(S_d)_{ir}/PGL_r(\C)$ be the quotient map.
The phase space of the Beauville system is 
$ 
    \cM(r,d)=M_r(S_d)_{ir}/PGL_r(\mathbb{C})
$.
It was shown in \cite{Beauville90} that 
if $P\in V(r,d)$ def\/ines a smooth spectral curve $C_P$,
then $\psi^{-1}(P)/PGL_r(\C)$ is isomorphic to 
the complement of the theta divisor in $\mathrm{Pic}^{g-1}(C_P)$,
where $g = \frac{1}{2}(r-1)(rd-2)$ is the genus of $C_P$.

Def\/ine the vector f\/ields $Y_i^{(k)}$ on $M_r(S_d)$  by
\begin{gather}\label{eq:Y}
   \sum_{i=0}^{dk-1}a^iY_i^{(k)}(A(x))=\frac{1}{x-a}[A(x),A(a)^k],
  \qquad \text{for} \quad k=1,\ldots, r-1.
\end{gather} 
Here we have identif\/ied the tangent space at each $A(x)\in M_r(S_d)$ with 
$M_r(S_d)$.
In \cite{Beauville90}, it was shown that 
$\eta_*Y_i^{(k)}$ 
generate the $g$-dimensional space of translation invariant vector f\/ields
on $\mathrm{Pic}^{g-1}(C_P)$.

\subsection{Poisson structure}

We use the following shorthand notations:
\[
  \Mbar=M_r(S_{d+1}),\quad 
  M=M_r(S_d)          ,\quad 
  M_{ir}=M_r(S_d)_{ir},\quad
  \cM=\cM(r,d),       ,\quad 
 G=PGL_r(\C).
\]
Extending the result of \cite[\S~5]{Beauville90},
we are to equip $\cM$ with a family of compatible Poisson structures
depending on 
a polynomial $\phi(x) \in S_{d+2}$:
 \begin{gather*}
   \phi(x) = \sum_{i=0}^{d+2}\sigma_{i}x^i.
 \end{gather*}

Note that 
a Poisson structure on $\cM$ is equivalent to
a Poisson algebra structure on the sheaf of $G$-invariant functions $\mathcal{O}_{M_{ir}}^G$
on $M_{ir}$.
Moreover, 
a Poisson algebra structure on $\mathcal{O}_M^G$ induces  
that on $\mathcal{O}_{M_{ir}}^G$ since
$M_{ir}$ is an open subset of $M$.

Consider the following Poisson structure on  $\Mbar$: 
\begin{gather}
  \label{Poisson-basic}
  \{A_{ij}(x) , A_{kl}(y) \}
  =  
  \delta_{i,l} \frac{A_{kj}(x)\phi(y)-\phi(x)A_{kj}(y)}{x-y}
  -\delta_{k,j}\frac{A_{il}(x)\phi(y)-\phi(x)A_{il}(y)}{x-y}.
\end{gather}
Let $\iota:M \hookrightarrow \Mbar$ be the closed immersion. 
Let $\mathcal{I}_M$ be the ideal sheaf of $\iota$.
Writing $\alpha$ for the natural projection 
$\mathcal{O}_{M^\bullet} \to 
\mathcal{O}_{M^\bullet} / \mathcal{I}_M = \iota_\ast \mathcal{O}_M$,
we set 
$\mathcal{N} := \alpha^{-1}(\iota_\ast \mathcal{O}_M^G) 
\subset \mathcal{O}_{M^\bullet}$.

\begin{proposition}
  \label{prop:Poisson-Beauville}
  $\mathcal{N}$ is a Poisson subalgebra of $\mathcal{O}_{\Mbar}$.
  This Poisson algebra structure induces that  on 
  $\mathcal{N}/\mathcal{N}\cap \mathcal{I}_M$, hence on $\mathcal{O}_{M}^G$.
\end{proposition}

\begin{proof}
Let $U$ be an open subset of $\Mbar$.
Let us write $F,H\in \mathcal{N}(U)$ as
 \begin{gather*}
    F = f+\sum_{1\leq i,j\leq r}A_{ij;d+1} f_{ij},
    \qquad
 H = h+\sum_{1\leq i,j\leq r}A_{ij;d+1} h_{ij},
 \end{gather*}
with $f,h\in \mathcal{O}_{\Mbar}(U)^{PGL_r(\C)}$
and $f_{ij},h_{ij}\in \mathcal{O}_{\Mbar}(U)$.
Note that the Hamiltonian vector f\/ields of $A_{ij;d+1}$ ($1\leq i,j\leq r$) 
are proportional to 
the vector f\/ields $X_{[E_{ji}]}$ $(1\leq i,j\leq r)$ 
which generate  the inf\/initesimal actions
corresponding to $[E_{ji}]\in \Lie PGL_r(\C)$:
 \begin{gather*}
     \{A_{ij;d+1},* \} = -\sigma_{d+2}X_{[E_{ji}]} \qquad (1\leq i,j\leq r).
 \end{gather*}
Note also that
$PGL_r(\C)$-invariant functions $f,h$ 
vanish when one applies $X_{[E_{ji}]}$ as  derivations.
Combining these facts, we obtain
\[
       \{F,H\} = \{f,h\} + \sum_{1\leq i,j\leq r}A_{ij;d+1} l_{ij},
\]
where $l_{ij} \in \mathcal{O}_{\Mbar}(U)$.
Since \eqref{Poisson-basic} is $G$-invariant, 
$\{f,h\}\in \mathcal{O}_{\Mbar}^{G}$.
Thus we have
\[
        \{F,H\} \in \mathcal{N}(U),  \text{ and } 
       \alpha\bigl(\{F,H\}\bigr) = \alpha\bigl(\{f,h \}\bigr).
\]
Therefore $\mathcal{N}$ is a subalgebra of $\mathcal{O}_{\Mbar}$ with respect to \eqref{Poisson-basic} 
and this Poisson algebra structure
induces a Poisson algebra structure 
on $\mathcal{N}/\mathcal{N}\cap \mathcal{I}_M$.
\end{proof}

\begin{remark}
The Poisson structure constructed in \cite{Beauville90}
corresponds to the case when 
$\phi(x)$ is monic of  degree $d+2$ and has only simple roots.
With such $\phi(x)$,
the Poisson structure \eqref{Poisson-basic} on $M_r(S_{d+1})$
is equivalent to the canonical Poisson structure
on $M_r(\C)^{d+2}$, on which the discussion in \cite{Beauville90} is based.   
See Appendix \ref{section:canonical-Poisson-structure}
for the explicit correspondence.
\end{remark}

\subsection{Multi-Hamiltonian structure}\label{sec:Beauville-m-Ham}

We def\/ine a family of Poisson structures on $\cM$:
\begin{definition}
  \label{def:m-poisson-B}
  For $\phi(x)\in S_{d+2}$, 
  $\{\cdot,\cdot\}_{\phi}:\mathcal{O}_{\cM}\times\mathcal{O}_{\cM} 
                     \to \mathcal{O}_{\cM}$ 
  denotes  the Poisson structure on~$\cM$ obtained in 
  Proposition \ref{prop:Poisson-Beauville}.
  For $0\leq i\leq d+2$,
  we write $\{\cdot,\cdot\}_i:=\{\cdot,\cdot\}_{\phi}$ with $\phi(x)=x^i$.
\end{definition}

By construction, 
the Poisson structures are compatible:
\[
  \{\cdot,\cdot \}_{c_1\phi_1+c_2\phi_2}
  =
  c_1\{\cdot,\cdot \}_{\phi_1}+c_2\{\cdot,\cdot \}_{\phi_2}
  \qquad (c_1,c_2\in \C,\phi_1(x),\phi_2(x)\in S_{d+2}).
\]

Def\/ine the $PGL_r(\C)$-invariant functions 
$H^{(k)}_{i}$ ($1\leq k\leq r$, $0\leq i\leq kd$) on $M_r(S_d)$ by
\begin{gather*}
   \frac{1}{k} \, \Tr\, A(x)^k = \sum_{i=0}^{kd} H^{(k)}_{i} x^i \qquad
   \text{for} \quad A(x) \in M_r(S_d).
\end{gather*}

\begin{lemma}\label{lemma:Ham-Y}  
{\rm (Cf.~\cite[proposition in (5.2)]{Beauville90}.)}
The Hamiltonian vector field of $H_j^{(k)}$ 
  $(1\leq k\leq r, $ $0\leq j\leq dk)$ with respect to the Poisson structure 
  $\{\cdot,\cdot\}_{\phi}$ is related to the vector fields \eqref{eq:Y} as follows:
\[
  \{ H_j^{(k)},*\}_{\phi} 
  =\sum_{i=0}^{\mathrm{min}(j,d+2)}\sigma_i  \,\eta_*Y_{j-i}^{(k-1)}.
\]
In particular,  $H_j^{(1)}$ $(0\leq j\leq d)$ are Casimir functions.
\end{lemma}

\begin{proof}
By direct calculation, we can show that
for each $\phi(x) \in S_{d+2}$ and $k \geq 1$, 
the Hamiltonian vector f\/ield of 
$\frac{1}{k}\,\Tr A(a)^k$ $(a\in \mathbb{P}^1)$ on $M_r(S_{d+1})$ 
with respect to the Poisson structure \eqref{Poisson-basic} is 
 \begin{gather*}
    \frac{\phi(a)}{x-a}[A(x),A(a)^{k-1}]. 
 \end{gather*}
It is easy to show that this is tangent to $M_r(S_d)$ and that
its restriction to $M_r(S_d)$ is 
\begin{gather}
  \label{vec-M_d+1} 
    \frac{\phi(a)}{x-a}[A(x),A(a)^{k-1}]
  = \sum_{i=0}^{d+2}\sum_{j=0}^{d(k-1)-1} \sigma_i a^{i+j}Y_j^{(k-1)}(A(x)).
\end{gather}
By Proposition \ref{prop:Poisson-Beauville},
the corresponding Hamiltonian vector f\/ield 
is given by a push forward of \eqref{vec-M_d+1} by $\eta$.
Comparing the coef\/f\/icients of powers of $a$, we obtain the lemma.
\end{proof}

\begin{theorem}\label{thm:m-Ham-Beauville}
  (i)
  Each $\eta_*Y_j^{(k)}$ is 
  a multi-Hamiltonian vector field  with respect to
  the Poisson structures $\{\cdot,\cdot\}_i$ $(i=0,\ldots,d+2)$: 
  \begin{gather*}
  \eta_*Y_j^{(k)} = \{H^{(k+1)}_j , \ast \}_0
  = \{H^{(k+1)}_{j+i} , \ast \}_i
  \end{gather*}     
  for $1\leq k \leq r-1$ and $0\leq j\leq kd-2$. 
  
  (ii)  With respect to $\{\cdot,\cdot\}_i$ $(0\leq i\leq d+2)$,  
  $H_0^{(k)}\!,\ldots, H_{i-1}^{(k)}\!$ and 
  $H_{d(k-1)+i-1}^{(k)},\ldots, H_{dk}^{(k)}\!$   $(1\leq k\leq r)$  
  are Casimir functions.
\end{theorem}

\begin{proof} 
By Lemma \ref{lemma:Ham-Y}, we obtain
\begin{alignat*}{3}
  & \{ H_j^{(k+1)},\ast \}_i =\eta_* Y_{j-i}^{(k)}\qquad
  && \text{for}\quad i\leq j\leq dk+i-1,&
  \\
  & \{H_j^{(k+1)},*\}_{i} = 0\qquad
  && \text{for}\quad   0\leq j\leq i-1 \quad \text{and}\quad  dk+i\leq j \leq  d(k+1).&
\end{alignat*}
Moreover,
$\eta_{*}Y_{dk-1}^{(k)}=0$ since $Y_{dk-1}^{(k)}$ 
is tangent to $PGL_r(\C)$-orbits by the def\/inition \eqref{eq:Y}.
This proves the theorem.
\end{proof}

\subsection[Poisson structure for representatives of ${\mathcal M}(2,d)$]{Poisson structure for representatives of 
$\boldsymbol{\cM(2,d)}$} 

In this subsection, 
we explicitly write down  the Poisson structure $\{\cdot,\cdot\}_{\phi}$
in the case of $r=2$.
We also discuss how this is related to the known Poisson structures
on the Mumford system.

Consider the subspace $\mathcal{S}_{\infty}\subset M_2(S_d)$ def\/ined by
\begin{gather*}
\mathcal{S}_{\infty}=\Biggl\{  
  S(x)   
  = 
  \begin{pmatrix}
  v_d & 0 \\
  1 & 0 
  \end{pmatrix} x^d\!
  +\!
  \begin{pmatrix}
  v_{d-1} & u_{d-1} \\
  w_{d-1} & 0 
  \end{pmatrix} x^{d-1}\!
  + \!
  \begin{pmatrix}
  v_{d-2} & u_{d-2} \\
  w_{d-2} & t_{d-2} 
  \end{pmatrix} x^{d-2}\!
  + \cdots
  \Bigg|\,
          u_{d-1}\neq 0
\Biggr\}.
\end{gather*}
In Section~\ref{section:rep-Beauville}, we will see that
$\mathcal{S}_{\infty}$ is a space of representatives for $M_{\infty}$
which is an open subset of 
\[
  M_{2d}=\big\{A(x)\in M_2(S_d)\mid H_{2d}^{(2)}=0\big\}.
\]

\begin{lemma}\label{lem:Poisson-Sinfty2}
  If $\phi(x) \in S_{d+1}$,
  \eqref{Poisson-basic} induces a Poisson structure on $\mathcal{S}_{\infty}$.
\end{lemma}

\begin{proof}
In this proof we write $M$ for $M_2(S_d)$.
By Proposition \ref{prop:Poisson-Beauville}, 
we have the Poisson algebra structure $\{ \cdot,\cdot\}_\phi$ 
on the sheaf $\mathcal{O}_{M}^G$.
Moreover, 
$H_{2d}^{(2)}$ is its Casimir function since $\deg\phi\leq d+1$ 
(Theorem \ref{thm:m-Ham-Beauville}).
Therefore the Poisson algebra structure induces that  on
$\mathcal{O}_{M}^G/\mathcal{O}_M^G\cap \mathcal{I}_{M_{2d}}$,
where $\mathcal{I}_{M_{2d}}$ is the ideal sheaf of $M_{2d}$ in $M=M_2(S_d)$.
Thus \eqref{Poisson-basic} induces the Poisson structure on 
$M_{\infty}/G\cong \mathcal{S}_{\infty}$.
\end{proof}

By a direct calculation 
(cf. proof of Proposition \ref{prop:Poisson-BV-2}),
we obtain the next result.
\begin{proposition}
  \label{prop:Poisson-B-2}
  For $\phi(x) = \sigma_{d+1} x^{d+1} + \cdots + \sigma_0 \in S_{d+1}$,
  the Poisson structure $\{\cdot,\cdot\}_{\phi}$ is written as follows
  \begin{gather}
    \{ S(x) \stackrel{\otimes}{,} S(y) \}_{\phi}
    =
    \phi(y) [r(x,y) , S(x) \otimes \mathbb{I}_2 ]
    - \phi(x) [\bar{r}(x,y), \mathbb{I}_2 \otimes S(y)]\nonumber    \\
\phantom{ \{ S(x) \stackrel{\otimes}{,} S(y) \}_{\phi}
    =}{}+ [K(x,y) , S(x) \otimes \mathbb{I}_2 ]
    - [\bar{K}(x,y) , \mathbb{I}_2 \otimes S(y)],\label{Poisson-S-infty}
  \end{gather}
where
\[
 \{ S(x) \stackrel{\otimes}{,} S(y) \}_{\phi}
 =
 \sum_{1\leq i,j,k,l \leq r} E_{ij} \otimes E_{kl} \,
                             \{S_{ij}(x) , S_{kl}(y)\}_{\phi}
\]
and
  \begin{gather*}
  r(x,y) = \frac{1}{x-y}\mathbb{P}_2 
            + 
            \frac{1}{u_{d-1}}
            \begin{pmatrix} v_d & 0 \\ 1 & 0 \end{pmatrix}
            \otimes
            \begin{pmatrix} 0 & 0 \\ 1 & 0 \end{pmatrix},
  \\
  \bar{r}(x,y) = \mathbb{P}_2 \cdot r(y,x) \cdot \mathbb{P}_2,
  \\
  K(x,y) = \frac{1}{u_{d-1}}
            \begin{pmatrix} 
              y v_d \sigma_{d+1} & - u_{d-1} \sigma_{d+1} \\ 
              -(w_{d-1} - y) \sigma_{d+1} +\sigma_d & 0 
            \end{pmatrix}
            \otimes
            \begin{pmatrix} 0 & S_{12}(y) \\ -S_{21}(y) & 0 \end{pmatrix},
  \\
  \bar{K}(x,y) = \mathbb{P}_2 \cdot K(y,x) \cdot \mathbb{P}_2,
  \qquad
  \mathbb{P}_2 = \sum_{1 \leq i,j \leq 2} E_{ij} \otimes E_{ji}.
  \end{gather*}
\end{proposition}
We write $F_j^{(1)}$ $(j=0,\ldots,d-2)$
for the vector field on $\mathcal{S}_\infty$ induced by $\eta_\ast Y_j^{(1)}$. 
As a consequence of Theorem \ref{thm:m-Ham-Beauville} and Proposition 
\ref{prop:Poisson-B-2} we obtain 
\begin{corollary}
  Each $F_j^{(1)}$ $(j=0,\ldots,d-2)$ is the multi-Hamiltonian vector field
  with respect to the Poisson structure \eqref{Poisson-S-infty}.
  They are written as the Lax form:
  \begin{gather}
    \sum_{j=0}^{d-2} y^j F_j^{(1)} \big( S(x)\big) 
    = \frac{1}{y^i}\big\{ H^{(2)}(y) , S(x) \big\}_i   
      = \left[ S(x) , \frac{1}{x-y} S(y) 
                    + \frac{S_{12}(y)}{u_{d-1}} 
                      \begin{pmatrix} v_d & 0 \\ 1 & 0 \end{pmatrix}
       \right],\label{vf-Sinf}
  \end{gather}
  for $i=0,\ldots,d+1$.
\end{corollary}

Now we derive a Poisson structure of the Mumford system from 
\eqref{Poisson-S-infty}.
The phase space $\mathcal{S}_{\rm Mum}$ of the Mumford system is
the subspace of $\mathcal{S}_\infty$ def\/ined as   
\[
     \mathcal{S}_{\rm Mum} =  \big\{S(x)\in\mathcal{S}_{\infty} \mid 
                            \Tr\, S(x)=0, u_{d-1}= 1\big\}.
\]

\begin{lemma}
  \eqref{Poisson-S-infty} induces a Poisson structure  on 
  $\mathcal{S}_{\rm Mum}$ if $\sigma_{d+1}=0$.
\end{lemma}

\begin{proof}
When  $\deg\phi(x)\leq d$, $H_{2d-1}^{(2)} = u_{d-1}$ 
is a  Casimir of $\{\cdot,\cdot\}_{\phi}$ by Theorem \ref{thm:m-Ham-Beauville}.
Therefore \eqref{Poisson-S-infty} induces a Poisson structure on 
$\mathcal{S}_{\rm Mum}$.
\end{proof}
This is the same as the Poisson structure in \cite[\S~5.1]{PenVan98}.
The Poisson structures in \cite[(4)]{FernadesVanhae01} 
and~\cite{SmirnovNakayashiki00} 
correspond to the case $\sigma_{d+1}=\sigma_{d}=0$ 
and the case $\phi(x)=x$
respectively.
The formula \eqref{vf-Sinf} reduces to the Lax form
for the Mumford system \cite[(7)]{FernadesVanhae01}.

\section[Multi-Hamiltonian structure on the Beauville-Vanhaecke system]{Multi-Hamiltonian 
structure\\ on the Beauville--Vanhaecke system}
\label{section:BV}

\subsection[The Beauville-Vanhaecke system]{The Beauville--Vanhaecke system}

Following \cite{IKY06}, we def\/ine the set $M'(r,d)$ and the group $G_r$ as 
\begin{gather*}
  M'(r,d) = 
  \left\{
  A(x) \in M_r(\C[x]) ~\Bigg|
  \begin{array}{c}
    A(x)_{11} \in S_d, ~~ A(x)_{1j}\in S_{d+1}, \\
    A(x)_{i1}\in S_{d-1}, ~~ A(x)_{ij} \in S_d,  
  \end{array}
    ~(2 \leq i,j \leq r)
  \right\},
\\
   G_r=\Bigg\{
             g(x)=
             \begin{pmatrix}
                1  & ^t \vec{b}_1x +\, ^t \vec{b}_0\\
                0  & B
             \end{pmatrix}
            \Bigg|\ 
            B \in GL_{r-1}(\C),\quad 
            \vec{b}_1,\vec{b}_0  \in\C^{r-1}
       \Bigg\}.
\end{gather*}
Here we use the notation such as
$\vec{b}$ for  a column vector and ${}^t \vec{b}$
for a row vector.
The group $G_r$ acts on $M'(r,d)$ by conjugation.
Let $\psi:M'(r,d)\to V(r,d)$ be the map
$\psi(A(x))=\det (y\mathbb{I}-A(x))$
and def\/ine
\[ 
  M'(r,d)_{ir}=\{A(x)\in M'(r,d) \mid  \text{the spectral curve 
  $C_{\psi(A(x))}$ is irreducible}\} .
\]
The $G_r$-action is free on $M_r(S_d)_{ir}$ \cite[Lemma 2.6]{IKY06}.
Let  $\eta':M'(r,d)_{ir}\to M'(r,d)_{ir}/G_r$ be the quotient map. 
The phase space of the Beauville--Vanhaecke system is
$ 
    \cM'(r,d)=M'(r,d)_{ir}/G_r
$.
It was shown that if $P\in V(r,d)$ def\/ines a smooth spectral curve $C_P$,
then $\psi^{-1}(P)/G_r$ is isomorphic to the complement of the intersection of 
$r$-translates of the theta 
divisor in $\textrm{Pic}^g(C_P)$ \cite[Theorem 2.8]{IKY06}.

Def\/ine the vector f\/ields $Y_i^{(k)}$ on $M'(r,d)$  by
\begin{gather} 
  \label{Y-BV}
   \sum_{i=0}^{kd} a^iY_i^{(k)}(A(x))=\frac{1}{x-a}[A(x),A(a)^k],
   \qquad \text{for}\quad k=1,\ldots, r-1.
\end{gather} 
It was shown that 
$\eta'_* Y_i^{(k)}$  
generate
the $g$-dimensional space of translation invariant vector f\/ields
on $\textrm{Pic}^g(C_P)$\footnote{Although $Y_i^{(k)}$ is not $G_r$-invariant, $\eta'_* Y_i^{(k)}$ 
is well-def\/ined 
because the dif\/ference between $g(x)_*Y_i^{(k)}$ and $Y_i^{(k)}$
is tangent to $G_r$-orbits \cite[Lemma 3.2]{IKY06}.}.

\subsection{Poisson structure}

We  equip $\cM'(r,d)$ with a family of Poisson structures,
extending the results in \cite[\S~3]{IKY06}.
The key idea is that
\eqref{Poisson-basic} induces the Poisson structure on $\cM'(r,d)$
as  in the case of the Beauville system.
However, due to the technical dif\/f\/iculties arising from the $G_r$-action,
we need a modif\/ication of the argument. 

We use the following shorthand notations:
\[
  M'=M'(r,d),\qquad M'_{ir}=M'(r,d)_{ir},\qquad
  \cM'=\cM'(r,d).
\]
Let us write $A(x)\in \Mbar$ and $A(x)\in M'$ as
\begin{gather*}
  A(x)=
  \begin{pmatrix}
  v(x) & ^t \vec{w}(x)\\ 
  \vec{u}(x) & t(x)
  \end{pmatrix},
\end{gather*}
where 
 \begin{gather*}
   v(x)=A_{11}(x), \qquad ^t \vec{w}(x)=(A_{12}(x),\ldots,A_{1r}(x)),\\
  \vec{u}(x)= ^t (A_{21}(x),\ldots,A_{r1}(x)),    
              \qquad  t(x)=\bigl(A_{ij}(x)\bigr)_{2\leq i,j\leq r}.
 \end{gather*}
Let $\iota':M'\hookrightarrow \Mbar$ be the closed immersion and 
$\pi':\Mbar\to M'$ be the surjection:
\[
      A(x) =
         \begin{pmatrix}
           \sum\limits_{k=0}^{d+1} v_k x^k &  \sum\limits_{k=0}^{d+1} {}^t\vec{w}_k x^k \vspace{1mm}\\ 
           \sum\limits_{k=0}^{d+1}\vec{u}_k x^k  & \sum\limits_{k=0}^{d+1} t_k x^k
        \end{pmatrix}     
      \mapsto 
         \begin{pmatrix}
           \sum\limits_{k=0}^d v_k x^k     &  \sum\limits_{k=0}^{d+1} {}^t\vec{w}_k x^k \vspace{1mm}\\ 
           \sum\limits_{k=0}^{d-1}\vec{u}_k x^k  &  \sum\limits_{k=0}^d t_k x^k
        \end{pmatrix} .  
\]
Note that $\pi'\circ \iota'=id_{M'}$.
Let $\gamma$ be the composition of the morphisms: 
\[
   \Hom((\mathcal{O}_{\Mbar})^2,\mathcal{O}_{\Mbar})
   \stackrel{\pi_\ast^\prime}{\to}
   \Hom((\pi'_*\mathcal{O}_{\Mbar})^2,\pi'_*\mathcal{O}_{\Mbar})\to
   \Hom((\mathcal{O}_{M'})^2,\mathcal{O}_{M'}),
\]
where the second morphism is given by
\[
\Phi\mapsto \bigl[(\mathcal{O}_{M'})^2\stackrel{(\pi^{\prime \#})^2}{\to}
             (\pi'_*\mathcal{O}_{\Mbar})^2 \stackrel{\Phi}{\to}
            \pi'_* \mathcal{O}_{\Mbar} \stackrel{\pi'_*(\iota^{\prime \#})}{\to}    
           \pi'_* \iota'_*\mathcal{O}_{M'}=O_{M'} \bigr].
\]

We def\/ine $\{\cdot,\cdot\}^{BV} \in 
\Hom ((\mathcal{O}_{M'})^2, \mathcal{O}_{M'})$ 
to be the image of \eqref{Poisson-basic} 
by $\gamma$.
For the coordinate functions $A_{ij;k}$ of $ M'$,
it is written explicitly as
\begin{gather}\label{Poisson-basic-BV}
\{A_{ij}(x),A_{kl}(y)\}^{BV}\!=\!
\!\Bigg[\!
  \delta_{i,l} \frac{A_{kj}(x)\phi(y)-\phi(x)A_{kj}(y)}{x-y}
  -\delta_{k,j}\frac{A_{il}(x)\phi(y)-\phi(x)A_{il}(y)}{x-y}
\!\Bigg]_{\!\leq d_{ij},\leq d_{kl}}\hspace{-20mm}
\end{gather}
where $[\cdot]_{\leq d_{ij},\leq d_{kl}}$ means 
 taking the terms whose degree in $x$ is  smaller or equal to $d_{ij}$
and whose degree in $y$ is smaller or equal to $d_{kl}$.
Here $d_{ij}=d$, $d_{1j}=d+1$, $d_{i1}=d-1$ for $2\leq i,j\leq r$ 
and $d_{11}=d$.

\begin{proposition}
  \label{lem:Poisson-BV}
  The sheaf $\mathcal{O}^{G_r}_{M'}$ of $G_r$-invariant regular functions
  on $M'$ is closed with respect to \eqref{Poisson-basic-BV}.
  Moreover, \eqref{Poisson-basic-BV} is a Poisson algebra structure on  
  $\mathcal{O}_{M'}^{G_r}$.
\end{proposition}
The proof is delegated to Subsection~\ref{section:proof}.
As an immediate consequence of this proposition, 
we obtain the Poisson algebra structure on $\mathcal{O}_{M'_{ir}}^{G_r}$,
which is equivalent to the Poisson structure on $\cM'$.

\begin{remark}
The Poisson structure constructed in \cite[\S~3]{IKY06}
corresponds to the case where  $\phi(x)$ is monic
of degree $d+2$ and has only simple roots.
\end{remark}

\subsection[Proof of Proposition 3]{Proof of Proposition \ref{lem:Poisson-BV}}
\label{section:proof}
We prove Proposition \ref{lem:Poisson-BV}
in the cases of $\deg\phi(x)=d+2$, 
$\deg\phi(x)= d+1$ and $\deg\phi(x)\leq d$ separately.

{\bf {The case of $\deg\phi(x)=d+2$}:}
We equip $\Mbar$ with the Poisson structure \eqref{Poisson-basic}.
We extend the $G_r$-action on $M'$ to $\Mbar$ as follows\footnote{In the case $\phi(x)$ has only simple roots,
this action is the same as the one used in \cite{IKY06}.
See Appendix \ref{section:Gr-action} for a~proof.}:
 \begin{gather}\label{Gr-action}
    G_r\ni g(x): A(x)\mapsto \tilde{A}(x),
 \end{gather}
where $\tilde{A}(x)\in M_r(S_{d+1})$ 
is the matrix uniquely determined by
 \begin{gather}
  \label{Gr-action-sub}
   g(x)^{-1}A(x)g(x)=\tilde{A}(x)+\phi(x)\hat{A}(x) ,
          \qquad \hat{A}(x)\in M_r(S_1).
 \end{gather}
By direct calculation, we can show that
the Poisson structure \eqref{Poisson-basic} is invariant with respect to  this $G_r$-action.

Let $\mathcal{I}_{M'}$ be the ideal sheaf of $\iota'$,
and set 
$\mathcal{N}' := {\alpha'}^{-1}(\iota'_\ast \mathcal{O}_{M'}^{G_r}) 
\subset \mathcal{O}_{M^\bullet}$
by writing $\alpha'$ for the natural projection 
$\mathcal{O}_{M^\bullet} \to 
\mathcal{O}_{M^\bullet} / \mathcal{I}_{M'} = \iota'_\ast \mathcal{O}_{M'}$.
\begin{lemma}\label{lem:d+2}
 (1) $\mathcal{N}'$ is a Poisson subalgebra of $\mathcal{O}_{\Mbar}$. 

 (2) The  Poisson algebra structure of (1) induces that  on 
     $\mathcal{N}'/\mathcal{N}'\cap\mathcal{I}_{M'}$, hence on $\mathcal{O}_{M'}^{G_r}$.
     Moreover, it is given by \eqref{Poisson-basic-BV}.
\end{lemma}

\begin{proof}
Note that for $f\in \mathcal{O}_{\Mbar}(U)$ where $U$ is any open subset,
\begin{gather*}
  \sum_{i=1}^r\{A_{ii;d+1},f\}=0,
  \\
  \{A_{ij;d+1},f\}=-\sigma_{d+2} X_{E_{ji}}f
  \qquad (2\leq i,j\leq r),
  \\
  \{A_{i1;d+1},f\}=-\sigma_{d+2} X_{E_{1i}}f,
  \qquad \{A_{i1;d},f\}=-\sigma_{d+2} X_{E_{1i}'}f-\sigma_{d+1}X_{E_{1i}}f 
  \qquad (2\leq i\leq r).
\end{gather*}
Here $X_{E_{ji}}$, $X_{E_{1i}}$, $X_{E_{1i}'}$ are 
the vector f\/ields generating the inf\/initesimal actions corresponding to 
$E_{ji}$, $E_{1i}$, $E_{1i}'=xE_{1i}\in \Lie G_r$:
 \begin{alignat*}{3}
   & X_{E_{ij}}(A(x))  = [A(x),E_{ij}]\quad &&(2\leq i,j\leq r),&\\
   & X_{E_{1j}}(A(x))  = [A(x),E_{1j}]\quad 
                              && (2\leq j\leq r),&\\
   & X_{E_{1j}'}(A(x)) = \left[x A(x)-
                          \frac{\sigma_{d+1}}{\sigma_{d+2}}A^{d+1},E_{1j}\right]\qquad 
                                    &&(2\leq j\leq r).
\end{alignat*}
By the same argument as that of Proposition \ref{prop:Poisson-Beauville},
we can show that
$\mathcal{N'}$ is a Poisson subalgebra of $\mathcal{O}_{M'}$
and that this induces a Poisson algebra structure on 
$\mathcal{N'}/\mathcal{N'}\cap I_{M'}$, hence on $\mathcal{O}_{M'}^{G_r}$.
By construction, the Poisson algebra structure on $\mathcal{O}_{M'}^{G_r}$
coincides with the restriction of \eqref{Poisson-basic-BV} to~$\mathcal{O}_{M'}^{G_r}$.
\end{proof}

{\bf The Case of $\deg\phi(x)= d+1$:}
For an open subset $U$ of $M'$, 
a function $F\in \mathcal{O}_{M'}^{G_r}(U)$ is cha\-rac\-terized 
by the condition 
$X_{E_{ij}}F=X_{E_{1j}}F=X_{E_{1j}'}F=0$ ($2\leq i,j\leq r$)
where $X_{E_{ij}}$, $X_{E_{1j}}$, $X_{E_{1j}'}$
are the inf\/initesimal action on $M'$corresponding to $E_{ij},E_{1j},E_{1j}'\in \Lie G_r$:
\begin{alignat}{3}
   & X_{E_{ij}}(A(x))  = [A(x),E_{ij}] \qquad && (2\leq i,j\leq r),&\nonumber\\
   & X_{E_{1j}}(A(x))  = [A(x),E_{1j}]
                              \qquad && (2\leq j\leq r),\label{actionG_r}& \\
   & X_{E_{1j}'}(A(x)) = [x A(x),E_{1j}]
                                    \qquad && (2\leq j\leq r).&\nonumber
\end{alignat}
Using this fact, we can show that $\mathcal{O}_{M'}^{G_r}$
is closed with respect to $\{\cdot,\cdot\}^{BV}$.
We can also show that the Jacobi identity  holds on~$\mathcal{O}_{M'}^{G_r}$
although it  does not on~$\mathcal{O}_{M'}$.
Thus $\{\cdot,\cdot\}^{BV}$ is a Poisson algebra structure on 
$\mathcal{O}_{M'}^{G_r}$.

{\bf The Case of $\deg\phi(x)\leq d$:}
When $\deg\phi(x)\leq d$, $\{\cdot,\cdot\}^{BV}$ 
satisf\/ies the Jacobi identity on~$\mathcal{O}_{M'}$
since it gives $\mathcal{O}_{M'}$ a Poisson subalgebra structure.
By using the action \eqref{actionG_r} of Lie $G_r$ on~$M'$, 
we can show that 
$\mathcal{O}_{M'}^{G_r}$ is closed with 
respect to $\{\cdot,\cdot\}^{BV}$ as in the case of $\deg \phi(x) = d+1$.
Therefore $\{\cdot,\cdot\}^{BV}$ is a Poisson algebra structure 
on $\mathcal{O}_{M'}^{G_r}$.

\subsection{Multi-Hamiltonian structure}

\begin{definition}
  \label{def:m-poisson-BV}
  For $\phi(x)\in S_{d+2}$, 
  $\{\cdot,\cdot\}'_{\phi}:\mathcal{O}_{\cM'}\times \mathcal{O}_{\cM'}\to 
  \mathcal{O}_{\cM'}$ denotes
  the Poisson structure 
  def\/ined in Proposition~\ref{lem:Poisson-BV}.
  For $0\leq i\leq d+2$,
  we write $\{\cdot,\cdot\}_i':=\{\cdot,\cdot\}_{\phi}'$ with $\phi(x)=x^i$.
\end{definition}

By construction, these  Poisson structures are compatible 
in the sense of Subsection~\ref{sec:Beauville-m-Ham}.

Def\/ine the $G_r$-invariant functions 
$H^{(k)}_{i}$ ($1\leq k\leq r$, $0\leq i\leq kd$) on $M'(r,d)$ by
\begin{gather*}\notag
   \frac{1}{k} \,\Tr\, A(x)^k = \sum_{i=0}^{kd} H^{(k)}_{i} x^i \qquad
   \text{for} \quad A(x) \in M'(r,d).
\end{gather*}

\begin{lemma}\label{lemma:Ham-Y-BV} {\rm (Cf. \cite[Lemma 3.10]{IKY06}.)}
  The Hamiltonian vector field 
  of $H_j^{(k)}$ $(1\leq k\leq r,0\leq i\leq dk)$ is 
  related to the vector fields \eqref{Y-BV} as 
\[
  \{ H_j^{(k)},\ast\}_{\phi}'
  =\sum_{i=0}^{\mathrm{min}(j,d+2)}\sigma_i  \eta_*Y_{j-i}^{(k-1)}.
\]
  In particular,  $H_j^{(1)}$ $(0\leq j\leq d)$ are Casimir functions of 
  $\{\cdot,\cdot\}_{\phi}'$.
\end{lemma}
This lemma can be proved in the same way as Lemma \ref{lemma:Ham-Y}
using Proposition \ref{lem:Poisson-BV}.
From Lemma~\ref{lemma:Ham-Y-BV}, we obtain the following theorem
similar to Theorem~\ref{thm:m-Ham-Beauville}.
\begin{theorem}\label{thm:m-Ham-BV}
  (i)
  Each $\eta_*Y_j^{(k)}$ is a  
  multi-Hamiltonian vector field  with respect to  
  the Poisson structures $\{\cdot,\cdot\}_i'$ $(i=0,\ldots,d+2)$:
  \begin{gather*}
    \eta_*Y_j^{(k)} =
      \{H_j^{k+1},\ast\}_0' =\{H_{j+i}^{k+1},\ast\}_i'
  \end{gather*}     
  for  $1\leq k \leq r-1$ and  $0\leq j\leq kd-2$.

  (ii)  With respect to $\{\cdot,\cdot\}_i'$ $(0\leq i\leq d+2)$,  
   $H_0^{(k)}\!,\ldots, H_{i-1}^{(k)}\!$  and 
   $H_{d(k-1)+i-1}^{(k)},\ldots, H_{dk}^{(k)}\!$ $(1\leq k\leq r)$ 
  are Casimir functions. 
\end{theorem}

\subsection[Poisson structure for representatives of ${\mathcal M}'(2,d)$]{Poisson structure for 
representatives of $\boldsymbol{{\mathcal M}'(2,d)}$}

We present the Poisson structure $\{\cdot,\cdot\}_{\phi}'$ in the case $r=2$
on the space of representatives $\mathcal{S}'_{\infty}$
of $\cM'(2,d)$ constructed in \cite[\S~4]{IKY06}.

The space $\mathcal{S}'_\infty$ is written as 
\begin{gather*}
  \mathcal{S}'_\infty\! 
  =
  \Bigg\{
  S(x)
      =
        \begin{pmatrix}
          0 & w_{d+1} \\ 
          0 & 0
        \end{pmatrix} x^{d+1}\!
        +\!
  \begin{pmatrix}
    v_d & w_d \\
    0 & 0
  \end{pmatrix} x^{d}\!
  +\!
  \begin{pmatrix}
    v_{d-1} & w_{d-1}\\
    1 & 0 \\
  \end{pmatrix} x^{d-1}\!
  +\text{lower terms in $x$}
  \Bigg\}.\!\!
\end{gather*}

We obtain the following result by a direct calculation.
\begin{proposition}
  \label{prop:Poisson-BV-2}
  For $\phi(x) = \sigma_{d+2} x^{d+2} + \sigma_{d+1} x^{d+1} + 
  \cdots + \sigma_0$,
  the Poisson bracket $\{\cdot,\cdot\}_{\phi}'$ is written as follows:
\begin{gather}
  \{ S(x) \stackrel{\otimes}{,} S(y) \}_{\phi}'  = 
  \phi(y) [  r(x,y) , S(x)\otimes \mathbb{I}_2 ] 
  - \phi(x) [ \bar{r}(x,y) , \mathbb{I}_2 \otimes S(y) ] 
 \nonumber \\
\phantom{\{ S(x) \stackrel{\otimes}{,} S(y) \}_{\phi}'  = }{}+ B(x,y) [ K(x,y) , S(x)\otimes \mathbb{I}_2 ] 
   - B(x,y) [  \bar{K}(x,y) , \mathbb{I}_2 \otimes S(y) ], \label{Poisson-S'inf}
  \end{gather}
where 
\begin{gather*}
  r(x,y) = \frac{1}{x-y} \mathbb{P}_2 
            + 
            \begin{pmatrix}
              v_d & A(x+y) \\
              0 & 0
            \end{pmatrix}
            \otimes
            \begin{pmatrix}
              0 & 1 \\ 0 & 0
            \end{pmatrix}, 
  \qquad
 \bar{r}(x,y) = \mathbb{P}_2 \cdot r(y,x) \cdot \mathbb{P}_2,
  \\
  K(x,y) = \begin{pmatrix}
              0 & B(x,y) \\ 0 & 0
            \end{pmatrix}
            \otimes 
            \begin{pmatrix}
              0 & -S_{12}(y) \\
              S_{21}(y) & 0
            \end{pmatrix},
  \qquad
  \bar{K}(x,y) = \mathbb{P}_2 \cdot K(y,x) \cdot \mathbb{P}_2,
  \\
 A(x) = w_{d+1}(x-u_{d-2}) + w_d,
  \\
  B(x,y) = \sigma_{d+2}(x^2 + y^2 + xy + (u_{d-2}-x-y) u_{d-2} - u_{d-3}) 
            + \sigma_{d+1}(u_{d-2}-x-y)  + \sigma_d.
\end{gather*}
\end{proposition}

\begin{proof}
In this proof, we write $A(x)\in M'$ as 
 \begin{gather*}
  A(x)=
   \begin{pmatrix}
    \sum\limits_{i=0}^{d} \tilde{v}_ix^i & \sum\limits_{i=0}^{d+1}\tilde{w}_i x^i\vspace{1mm}\\
    \sum\limits_{i=0}^{d-1}\tilde{u}_i x^i&\sum\limits_{i=0}^{d} \tilde{t}_i x^i
   \end{pmatrix}
\end{gather*}
and regard
 $\tilde{v}_i$ $(0\leq i\leq d)$, $\tilde{w}_i$ $(0\leq i\leq d+1)$,
 $\tilde{u}_i$ $(0\leq i\leq d-1)$, $\tilde{t}_i$ $(0\leq i\leq d)$
as coordinate functions of $M'$.
(Here we use $\tilde{\,}$ to distinguish
from 
$v_i$, $w_i$, $u_i$, $t_i$  which we use as coordinates of~$\mathcal{S}_{\infty}'$.)
The bracket $\{\cdot,\cdot\}^{BV}$ \eqref{Poisson-basic-BV} 
among $\tilde{v}_i,\ldots,\tilde{t}_i$
are explicitly written as follows:
 \begin{gather}
  \{\tilde{v}(x),\tilde{v}(y)\}=
  \{\tilde{u}(x),\tilde{u}(y)\}=
  \{\tilde{w}(x),\tilde{w}(y)\}=
  \{\tilde{t}(x),\tilde{t}(y)\}=
  \{\tilde{v}(x),\tilde{t}(y)\}=0,
  \nonumber\\
  \{\tilde{v}(x),\tilde{u}(y)\}=-
  \{\tilde{t}(x),\tilde{u}(y)\}=
  \frac{\tilde{u}(x)\phi(y)-\phi(x)\tilde{u}(y)}{x-y}
  \nonumber\\
\phantom{\{\tilde{v}(x),\tilde{u}(y)\}=}{} + \sigma_{d+2}\tilde{u}(y)x^{d+1}+\sigma_{d+2}\tilde{u}(x)y^{d+1}-
  (\sigma_{d+1}-\sigma_{d+2} x)\tilde{u}(x)y^{d},
  \nonumber\\
  \{\tilde{v}(x),\tilde{w}(y)\}=-
  \{\tilde{t}(x),\tilde{w}(y)\}=
  -\frac{\tilde{w}(x)\phi(y)-\phi(x)\tilde{w}(y)}{x-y}
  - \sigma_{d+2}\tilde{w}(y)x^{d+1},
  \nonumber\\
  \{\tilde{u}(x),\tilde{w}(y)\}=
  \frac{(\tilde{v}(x)-\tilde{t}(x))\phi(y)-
     \phi(x)(\tilde{v}(y)-\tilde{t}(y))}{x-y}
  \nonumber\\
\phantom{\{\tilde{u}(x),\tilde{w}(y)\}=}{}  +\sigma_{d+2}(\tilde{v}(y)-\tilde{t}(y))x^{d+1}-
  (\sigma_{d+1}-\sigma_{d+2}y)(\tilde{v}(y)-\tilde{t}(y))x^{d},\label{poisson-br}
\end{gather}
where 
$\tilde{v}(x):=\sum\limits_{i=0}^{d}\tilde{v}_ix^i$ and so on.

Similarly let
$v_i$ $( 0\leq i\leq d)$,
$w_i$ $(0\leq i\leq d+1)$,
$u_i$ $(0\leq i\leq d-2)$,
$t_i$ $(0\leq i\leq d-2)$ denote coordinate functions of 
$\mathcal{S}_{\infty}'$.
Let $(b_1,b_0,c)\in \mathbb{C}^2\times \mathbb{C}^*$ 
be the following coordinate functions of~$G_r$:
\begin{gather*}
G_r\ni \begin{pmatrix} 1& b_1x+b_0\\ 0& c 
\end{pmatrix}.
\end{gather*}
In the neighborhood of $\mathcal{S}_{\infty}'$,
$(v_i$ $(0\leq i\leq d)$, $w_i$ $(0\leq i\leq d+1)$,
$u_i$ $(0\leq i\leq d-2)$, $t_i$ $(0\leq i\leq d-2)$, $b_0$, $b_1$, $c)$ forms a local coordinate system of~$M'$. 
The transformation between the two coordinate systems is given by
\begin{gather}\label{coord-tr}
\begin{pmatrix}
\tilde{v}(x)&\tilde{w}(x)\\
\tilde{u}(x)&\tilde{t}(x)
\end{pmatrix}
=
\begin{pmatrix}
1&b_1x+b_0\\
0&c
\end{pmatrix}^{-1}
\begin{pmatrix}
v(x)&w(x)\\u(x)&t(x)
\end{pmatrix}
\begin{pmatrix}
1&b_1x+b_1\\
0&c
\end{pmatrix}.
\end{gather}
Substituting the RHS of \eqref{coord-tr} into \eqref{poisson-br}
and using the Leibniz rule, 
we obtain the system of equations for
brackets among $(v(x)$, $w(x)$, $u(x)$, $t(x)$, $b_0$, $b_1$, $c)$. Solving this
and restricting to~$\mathcal{S}_{\infty}'$ 
(i.e.\ setting $b_0=b_1=0$, $c=1$), 
we arrive at the result of Proposition \ref{prop:Poisson-BV-2}. 
\end{proof}

As in the case of the Beauville system,
we write $F_j^{(1)}$ $(j=0,\ldots,d-2)$ for
the vector f\/ield on $\mathcal{S}'_\infty$ induced by $\eta_\ast Y_j^{(1)}$.
From Theorem \ref{thm:m-Ham-BV} and Proposition \ref{prop:Poisson-BV-2}
we obtain
\begin{corollary}
  Each $F_j^{(1)}$ $(j=0,\ldots,d-2)$ is the multi-Hamiltonian vector field
  with respect to the Poisson structure \eqref{Poisson-S'inf}.
  They are written as the Lax form:
  \begin{gather*}
    \sum_{j=0}^{d-2} y^j F_j^{(1)} \big( S(x) \big)
    = \frac{1}{y^i}\{ H^{(2)}(y) , S(x) \}_i   
   \nonumber \\
\phantom{\sum_{j=0}^{d-2} y^j F_j^{(1)} \big( S(x) \big)}{}= \left[ S(x) , \frac{1}{x-y} S(y) 
                    + S_{21}(y)
                      \begin{pmatrix} 0 & A(x+y)\\ 0 & -v_d \end{pmatrix}
       \right],
  \end{gather*}
  for $i=0,\ldots,d+2$.
\end{corollary}

We remark that this Lax form already appeared in \cite[(4.9)]{IKY06}
for general $r$.

In closing this subsection,  
we discuss the Poisson structure on the even Mumford system.
The phase space of the even Mumford system is given by
\[
  \{S(x)\in \mathcal{S}'_{\infty} \mid  \Tr\, S(x)=0, w_{d+1}=1\}.
\]

\begin{lemma}
  If $\sigma_{d+2}=0$,
  \eqref{Poisson-S'inf} gives a Poisson structure on the phase space of 
  the even Mumford system.
\end{lemma}

\begin{proof}
By Theorem \ref{thm:m-Ham-BV},
$w_{d+1}=H^{(2)}_{2d}$ is Casimir of $\{\cdot,\cdot\}_{\phi}'$ 
if $\deg \phi(x) \leq d+1$.
Therefore in such a case, \eqref{Poisson-S'inf} 
induces a Poisson algebra structure on
$\mathcal{O}(\mathcal{S}'_\infty)/H_{2d}^{(2)} 
\mathcal{O}(\mathcal{S}'_\infty)$.
\end{proof}

The Poisson structure  in \cite{FernadesVanhae01} corresponds to
the case $\sigma_{d+2}=\sigma_{d+1}=\sigma_d=0$.

\section{Representatives of the Beauville system}
\label{section:rep-Beauville}

First we introduce some notations.
Let us def\/ine a subset $M_{\rm reg}$ of $M_r(\C)$:
\begin{gather*}
  M_{\rm reg} = \{A \in M_r(\C) \,|\, \deg(\text{the minimal polynomial of } A) = r
                            \}.
\end{gather*}
For $A\in M_r(\C)$, $A\in M_{\rm reg}$ is equivalent to the condition that only 
one Jordan block corresponds to each eigenvalue of $A$.
For $A\in M_{\rm reg}$, let $\alpha_1,\ldots,\alpha_k$ ($k<r$) be the distinct
eigenvalues and $\nu_1,\ldots,\nu_k$ be the size of the corresponding 
Jordan blocks. 
Def\/ine the subspace of $\C^r$ as 
\[
  W_{\alpha_i:j} = \{ \vec{u} \in \C^r \,|\, (A-\alpha_i \I_r)^j \vec{u} = 0 \} 
\]
for $0 \leq i \leq k$, $j \in \Z_{\geq 0}$.
The spaces $W_{\alpha_i:1}$ and $W_{\alpha_i:\nu_i}$ are 
respectively the eigenspace and the generalized eigenspace of $A$.
There is the f\/iltration 
\[
  W_{\alpha_i:\nu_i} \supset W_{\alpha_i:\nu_i-1} \supset
  \cdots \supset W_{\alpha_i:1} \supset W_{\alpha_i:0} = \{ \vec{0} \}.        
\]  
By the assumption of $A$, dim$(W_{\alpha_i:j} / W_{\alpha_i:j-1}) = 1$
for all $\alpha_i$ and $j=1,\ldots,\nu_i$.
We f\/ix a base $\vec{v}_{\alpha_i}(A)$ of $W_{\alpha_i:1}$.
Let $\Pi_{\alpha_i}$ be the projection map 
$\Pi_{\alpha_i} : \C^r \to W_{\alpha_i:\nu_i}$, and def\/ine 
\[
  V(A) = \{ \vec{u} \in \C^r \,|\, \Pi_{\alpha_i} (\vec{u}) 
                                 \notin W_{\alpha_i:\nu_i-1}
                                 \text{ for } i=0,\ldots,k \}.
\]

Now we introduce the subspaces $M_\infty$ and $\mathcal{S}_\infty$
of $M_r(S_d)$:
\begin{gather}
  M_\infty 
  =
  \Big\{ A(x) = \sum_{k=0}^d A_k x^k \in M_r(S_d) \,\Big|\, 
          A_d \in M_{\rm  reg}, \ \det A_d = 0, \
          A_{d-1} \vec{v}_0(A_d) \in V(A_d)     
  \Big\},
  \nonumber\\
  \mathcal{S}_\infty 
  = 
  \Big\{ A(x) \in M_r(S_d) \,|\, 
    A(x) = \omega x^d + \rho x^{d-1} + \text{lower terms in } x, 
    \ \omega \in \Omega,\ \rho \in \mathcal{T}
  \Big\},\label{sub-rep}
\end{gather}
where 
\begin{gather*}
\Omega=
  \Biggl\{
  \begin{pmatrix}
    -\beta_1 & \cdots & -\beta_{r-1} & 0\\
    1&0&\cdots & 0\\    
    \vdots& \ddots &\ddots&\vdots \\
    0&\cdots&1&0 
  \end{pmatrix}
  \in M_{r}(\mathbb{C})
  \,\Big|\, \beta_1,\cdots,\beta_{r-1} \in \C 
  \Biggr\}
  \\
\mathcal{T} =
  \{ \rho \in M_{r}(\mathbb{C}) \,|\,
  \rho_{1r} \neq 0, ~\rho_{jr} = 0 \text{ for } j=2,\ldots,r \}.
  \end{gather*}
The main result of this section is as follows:
\begin{proposition}
  \label{prop:rep}
  (i) $\mathcal{S}_\infty \subset M_\infty$.

  (ii)
  The action of $PGL_r(\C)$ on $M_\infty$ induces an isomorphism 
  $\mathcal{S}_{\infty} \times PGL_r(\C) \simeq M_{\infty}$.
  Thus the space $\mathcal{S}_\infty$ is a set of representatives of 
  $M_{\infty} / PGL_r(\C)$.
\end{proposition}

\begin{remark}
One can def\/ine $\cM_{c}$ and $\mathcal{S}_c$ for $c\in \C$ as
\begin{gather*}
  M_c
  =
  \Bigl\{ A(x) \in M_r(S_d) \,\Big|\, 
          A(c) \in M_{\rm reg}, \ \det A(c) = 0, \
          A'(c) \vec{v}_0(A(c)) \in V(A(c))                    
  \Bigr\},
  \\
  \mathcal{S}_c 
  = 
  \{ A(x) \in M_r(S_d) \, |\, 
     A(x) = \omega + \rho (x-c) + \text{higher terms in } (x-c), 
     ~\omega \in \Omega, \ \rho \in \mathcal{T}
  \Bigr\}.
\end{gather*}
Then Proposition \ref{prop:rep} also holds for $(M_c,\mathcal{S}_c)$.
\end{remark}

Let us recall the following lemmas on linear algebra.
\begin{lemma}  
  \label{lemma:A_d-1}
  Let  $A\in M_{\rm reg}$. For $\vec{u}\in\C^{r}$, the followings are equivalent

  (i) $\vec{u}\in V(A)$;

  (ii) ${}^t\!\vec{v}_{\alpha}(^t\!A)\cdot\vec{u}\neq 0$
  for all eigenvalues $\alpha$ of $A$;

  (iii) 
  the vectors $\vec{u}, A\vec{u}, \ldots, A^{r-1}\vec{u}$  
  generate $\C^r$.
\end{lemma}
The proof is left for readers.

For $A\in M_{\rm reg}$, set  
\[
\xi_i(A) = A^i + \beta_1(A) A^{i-1} + \cdots + \beta_i(A) \I_r \in M_r(\C)
  \qquad (i=1,\ldots,r-1),
\]
where $\beta_1(A),\ldots,\beta_r(A)$ 
are coef\/f\/icients of the characteristic polynomial of $A$,
$\det (y\mathbb{I}_r-A) = y^r + \beta_1(A) y^{r-1} + \cdots + \beta_r(A)$.

\begin{lemma}
  \label{lemma:g-A_d}   
  Let $A\in M_{\rm reg}$  and $\vec{u}\in V(A)$.

 (i) 
  The matrix
  $g(\vec{u},A) = (\vec{u},\xi_1(A) \vec{u},\ldots,\xi_{r-1}(A) \vec{u})
  \in M_r(\C)$ is invertible. Moreover, it satisfies
\[
  g(\vec{u},A)^{-1} A   g(\vec{u},A) 
  = \begin{pmatrix}
    -\beta_1(A) & \cdots & -\beta_{r-1}(A) & -\beta_r(A)\\
    1&0&\cdots & 0\\    
    \vdots& \ddots &\ddots&\vdots \\
    0&\cdots&1&0 
  \end{pmatrix}.
\] 

  (ii)
  $g(\vec{u},A)$ makes $B\in M_r(\C)$ into the following form
\[
  g(\vec{u},A)^{-1}B g(\vec{u},A)=
  \begin{pmatrix}
    \ast & \cdots & \ast & \ast\\
    \ast & \cdots & \ast & 0\\    
    \vdots& \vdots & \vdots & \vdots \\
    \ast & \cdots & \ast & 0
  \end{pmatrix} ,
\]
   if and only if $\vec{u}$ is an eigenvector of $B\xi_{r-1}(A)$.
   Moreover the $(1,r)$-th entry of the RHS is equal to the eigenvalue.
\end{lemma}

\begin{proof}
  (i) The invertibility of $g$ follows from Lemma \ref{lemma:A_d-1}. 
  Another claim is checked by a direct computation.

  (ii) Let $\tilde{B}=g(\vec{u},A)^{-1}Bg(\vec{u},A)$. If $\tilde{B}$ has the form of 
     the RHS, we obtain     $ B\xi_{r-1}(A)\vec{u}=\tilde{B}_{1r}\vec{u}$ by
    comparing     the $r$-th columns of $Bg$ and $g\tilde{B}$.
    Conversely
    if $\vec{u}$ is an eigenvalue of $B\xi_{r-1}(A)$, then we see by direct calculation that
    $\tilde{B}_{1r}$ is equal to its eigenvalue and 
    $\tilde{B}_{jr}=0$ for $2\leq j\leq r$.   
\end{proof}

\begin{lemma} 
  \label{lemma:xi}
  Let  $A\in M_{\rm reg},B\in M_r(\C)$ and assume that  $\det A = 0$. 

  (i)  
  $\xi_{r-1}(A) = c \vec{v}_0(A) \otimes {}^t \vec{v}_0(^tA)$
  for some $c \in \C^\times$.

  (ii) $B \vec{v}_0(A)$ is an eigenvector of  $ B \xi_{r-1}(A)$.
\end{lemma}
\begin{proof}
  (i) By the assumption on $A$, the rank of $\xi_{r-1}(A)$ is one, and  
  $\xi_{r-1}$ satisf\/ies $A \xi_{r-1}(A) = \xi_{r-1}(A) A = 0$.
  Thus $\xi_{r-1}(A)$ have to be written as 
  $c \vec{v}_0(A) \otimes {}^t \vec{v}_0(^t A)$ with some $c \in \C^\times$.

  (ii) By (i), any $\vec{w}\in \C^r$ satisfy
  $\xi_{r-1}(A)\vec{w} = c  \vec{v}_0(A)$ 
  with some $c \in \C^\times$. 
  By multiplying the both sides by $B$ from the left and 
  setting $\vec{w}=B\vec{v}_0(A)$,
  we see that $B \vec{v}_0(A)$ is an eigenvector of~$B \xi_{r-1}(A)$. 
\end{proof}

\begin{proof}[Proofs of Proposition \ref{prop:rep}] 
  (i)
  We write $S(x) = S_d x^d + S_{d-1} x^{d-1} + \cdots + S_0
  \in \mathcal{S}_\infty$ as
  \begin{gather}
    \label{concrete-S}
  S(x) 
  = 
  \begin{pmatrix}
    -\beta_1 & \cdots & -\beta_{r-1} & 0\\
    1&0&\cdots & 0\\    
    \vdots& \ddots &\ddots&\vdots \\
    0&\cdots&1&0 
  \end{pmatrix} x^d 
  +
  \begin{pmatrix}
    \ast & \cdots & \ast & \beta\\
    \ast & \cdots & \ast & 0\\    
    \vdots& \vdots & \vdots & \vdots \\
    \ast & \cdots & \ast & 0
  \end{pmatrix} x^{d-1} 
  + \text{lower terms in $x$}, 
  \end{gather}
  where $\beta \neq 0$. 
  Then it is easy to see $\det S_d = 0$,
  and we can set $\vec{v}_0(S_d) = {}^t(0, \ldots, 0,1)$. 
  A direct calculation shows that 
  $\vec{u} := S_{d-1} \vec{v}_0(S_d) = {}^t(\beta, 0, \ldots, 0)$
  and $S_d$ satisfy
\[
  \det(\vec{u}, S_d \vec{u}, \ldots, S_d^{r-1} \vec{u}) = 
  \det(\mathrm{diag}(\beta, \ldots, \beta)) = \beta^r \neq 0.
\]
  Thus we see $\vec{u} \in V(S_d)$ due to Lemma \ref{lemma:A_d-1},
  and the claim follows.

  (ii)
  It is easy to see that $M_\infty$ is invariant under the action of 
  $PGL_r(\C)$, thus the map 
\[
  \mu : \mathcal{S}_{\infty} \times PGL_r(\C) \to M_{\infty};
  (S(x), g) \mapsto g  S(x) g^{-1}
\] 
  is well-def\/ined.
  In the following we show that $\mu$ is bijection.

  First we show the surjectivity of $\mu$.
  For $A(x) = A_d x^d + A_{d-1} x^{d-1} + \cdots + A_0 \in M_\infty$, 
  set $g=g(A_{d-1} \vec{v}_0(A_d), A_d)$.
  By Lemma \ref{lemma:A_d-1}, $g$ is invertible
  and $g^{-1}A_d g\in \Omega$ by Lemma \ref{lemma:g-A_d}(i).
  By Lemma \ref{lemma:A_d-1} and \ref{lemma:xi}, 
  $A_{d-1} \vec{v}_0(A_d)$ is an eigenvector of  
  $A_{d-1}\xi_{r-1}(A_d)$ belonging to a nonzero eigenvalue. 
  Thus $ g^{-1} A_{d-1} g\in \mathcal{T}$ by Lemma \ref{lemma:g-A_d}(ii).
  Consequently we obtain $S(x) = g A(x) g^{-1} \in \mathcal{S}_\infty$,
  i.e.\ $\mu(S(x),g) = A(x)$.
  
  To check the injectivity of $\mu$, we only have to check 
  the following:
  for any $S(x)\in \mathcal{S}_{\infty}$, $g\in GL_r(\C)$ satisf\/ies
  $g^{-1}S(x)g\in \mathcal{S}_{\infty}$ 
  only when $g$ is a scalar matrix.    
  When $S(x)$ is given by~\eqref{concrete-S}, we get
\begin{gather*} 
  \xi _{r-1} (S_d)
  =
 {}^t(0,\ldots, 0,1)\cdot(1,\beta_1,\ldots,\beta_{r-1}),
  \\
  \big(\xi_k(S_d)\big)_{j1} = \delta_{k,j-1} \quad \text{for}\quad  k=1,\ldots,r-2.
\end{gather*}
The f\/irst equation implies that
 $\bigl(S_{d-1}\xi_{r-1}(S_d) \bigr)_{ij}=\delta_{i,1}\delta_{j,1}\beta$.
This matrix has only one nonzero eigenvalue $\beta$
and the corresponding eigenvector is $^t(a,0,\ldots,0)$ 
for some $a\in \C^{\times}$.
By Lemma~\ref{lemma:g-A_d}(ii), we only have to show that
$g(^t(a,0,\ldots,0),S_d) = c  \mathbb{I}_r$ for some $c \in \C^\times$.
This follows from the second equation.
\end{proof}

In closing, we give some remarks.
The space $M_{\infty}$ \eqref{sub-rep} is an af\/f\/ine subspace of 
\[
  \{ A(x) \in M_r(S_d) \,|\, H_{rd}^{(r)} = 0 \}
\]
which is the codimension one subspace of $\cM(r,d)$.
As a generalization of Lemma \ref{lem:Poisson-Sinfty2},
we easily obtain the following:
\begin{lemma}
  If $\phi(x) \in S_{d+1}$, then \eqref{Poisson-basic}
  induces the Poisson structure on $M_\infty / PGL_r(\C) \simeq \mathcal{S}_\infty$.
\end{lemma}
The space of representatives introduced by
Donagi and Markman \cite[Lemma 4.1]{DonagiMarkman96}
is a subspace of $\mathcal{S}_\infty$ \eqref{sub-rep}
def\/ined by
\[
  \{ S(x) = S_d x^d + \cdots + S_0 \in \mathcal{S}_\infty \,|\,  
       \Tr S(x) = 0, \  \beta_1(S_d) = \cdots = \beta_{r-1}(S_d) = 0 \}.
\]
The phase space discussed in \cite[\S~3.2]{SmirnovZeitlin02} is obtained
by removing the f\/irst condition in the above.

\appendix
\section{On the Poisson structure (\ref{Poisson-basic})}
\label{section:canonical-Poisson-structure}

Assume that $\phi(x)$ is a monic polynomial with only simple roots,
$a_1,\ldots,a_{d+2}$. 
Consider the following isomorphism $\varphi$ \cite[(5.6)]{Beauville90}:
\begin{gather}\label{map-varphi}
   \varphi: M_r(S_{d+1}) \to M_r(\mathbb{C})^{\oplus d+2}
         ,\
         A(x) \mapsto \bigl(c_1A(a_1),\ldots,c_{d+2}A(a_{d+2})\bigr),
\end{gather}
where $c_\alpha=\prod\limits_{\beta \neq \alpha}(a_\alpha-a_\beta)^{-1}$.
The inverse of 
$\varphi^{-1}$ is given by 
the Lagrange interpolation formula:
\[
   \varphi^{-1} :  M_r(\mathbb{C})^{d+2} \to M_r(S_{d+1}),\qquad
              \bigl(A^{(1)},\ldots, A^{(d+2)}\bigr) \mapsto 
                   \sum_{\alpha=1}^{d+2}A^{(\alpha)}
                   \prod_{\beta\neq \alpha}(x-a_{\beta}).
\]
Then the pullback by $\varphi$ of the canonical Poisson structure on 
$M_r(\mathbb{C})^{\oplus d+2}$:
\[
      \{A^{(\alpha)}_{ij},A_{kl}^{(\beta)}\} 
      = \delta_{\alpha,\beta}\,\bigl(\,\delta_{j,k}A_{il}^{(\alpha)}
        -\delta_{i,l}A_{kj}^{(\alpha)}\bigr)\qquad
        \text{for}\quad
        (A^{(1)},\ldots,A^{(d+2)})\in M_r(\mathbb{C})^{\oplus d+2},
\]
is equal to  \eqref{Poisson-basic}. 
This is easily checked if one uses the elementary identity
\[
A(x)\phi(y)-\phi(x)A(y)=(y-x)
\sum_{\alpha=1}^{d+2}A^{(\alpha)}\prod_{\mu\neq \alpha}(x-a_{\mu})(y-a_{\mu}).
\] 

\section[On the $G_r$-action (\ref{Gr-action})]{On the 
$\boldsymbol{G_r}$-action (\ref{Gr-action})}\label{section:Gr-action}

In the construction of Poisson structures in \cite{IKY06}, 
the isomorphism $\varphi$ given in \eqref{map-varphi} 
and the following $G_r$-action on $M_r(\mathbb{C})^{\oplus d+2}$ were used:
\begin{gather}\label{gr-action-ref}
  G_r \ni g(x): \bigl(A^{(1)},\ldots,A^{(d+2)}\bigr) \mapsto 
  \bigl(g(a_{\alpha})^{-1}A^{(\alpha)}\,
  g(a_{\alpha})\bigr)_{1\leq \alpha\leq d+2},
\end{gather}
where $a_{\alpha}\neq a_{\beta}$ if $\alpha\neq \beta$.
We show that this action is compatible with
the $G_r$-action \eqref{Gr-action} under the isomorphism $\varphi$
when $\phi(x)=\prod\limits_{\alpha=1}^{d+2}(x-a_{\alpha})$.

From \eqref{gr-action-ref}, we have 
\[
(\varphi^{-1}\circ g(x)\circ \varphi ) A(x)
=
\sum_{\alpha=1}^{d+2}c_{\alpha}\,g(a_{\alpha})^{-1}A(a_{\alpha})g(a_{\alpha})
             \prod_{\mu\neq\alpha}(x-a_{\mu}).
\] 
On the other hand, 
substituting $x=a_{\alpha}$ into \eqref{Gr-action-sub},
we have
$\tilde{A}(a_{\alpha})=g(a_{\alpha})^{-1}A(a_{\alpha})g(a_{\alpha})$.
Then expressing $\tilde{A}(x)$ by the Lagrange interpolation formula,
we see that 
$\tilde{A}(x)= (\varphi^{-1}\circ g(x)\circ \varphi ) A(x)$.

\subsection*{Acknowledgements}

The authors thank Takao Yamazaki for discussion and 
reading the manuscript.
Y.K. is a research fellow of the Japan Society for the Promotion of Science.

\pdfbookmark[1]{References}{ref}
\LastPageEnding

\end{document}